# Improved Multiple-Image-Based Reflection Removal Algorithm Using Deep Neural Networks

Tingtian Li, Yuk-Hee Chan, Daniel P. K. Lun

The Hong Kong Polytechnic University
Hong Kong SAR, China

# Improved Multiple-Image-Based Reflection Removal Algorithm Using Deep Neural Networks

Tingtian Li      Yuk-Hee Chan      Daniel P. K. Lun

The Hong Kong Polytechnic University
Hong Kong SAR, China
tingtianpolyu.li@connect.polyu.hk, {enyhchan, enpklun}@polyu.edu.hk

*Abstract*—**When imaging through a semi-reflective medium such as glass, the reflection of another scene can often be found in the captured images. It degrades the quality of the images and affects their subsequent analyses. In this paper, a novel deep neural network approach for solving the reflection problem in imaging is presented. Traditional reflection removal methods not only require long computation time for solving different optimization functions, their performance is also not guaranteed. As array cameras are readily available in nowadays imaging devices, we first suggest in this paper a multiple-image based depth estimation method using a convolutional neural network (CNN). The proposed network avoids the depth ambiguity problem due to the reflection in the image, and directly estimates the depths along the image edges. They are then used to classify the edges as belonging to the background or reflection. Since edges having similar depth values are error prone in the classification, they are removed from the reflection removal process. We suggest a generative adversarial network (GAN) to regenerate the removed background edges. Finally, the estimated background edge map is fed to another auto-encoder network to assist the extraction of the background from the original image. Experimental results show that the proposed reflection removal algorithm achieves superior performance both quantitatively and qualitatively as compared to the state-of-the-art methods. The proposed algorithm also shows much faster speed compared to the existing approaches using the traditional optimization methods.**

## I. INTRODUCTION

D IGITAL photography has become a daily activity for many people. While in every few months a new and more advanced digital camera is launched to the market, there are still problems in imaging that cannot be fully resolved with the existing imaging devices. One of them is reflection removal. It is common to capture images with reflection of unwanted scene in daily photography. The problem arises when taking pictures through a semi-reflective material, such as imaging the outside scenery through a window, or taking picture of underwater objects from above the water surface, etc. It is important to remove the unwanted reflection in the image since it does not only affect the visibility of the desired scene but also introduces ambiguity that perturbs the subsequent analysis on the image.

To solve the problem, traditionally photographers make use of a polarizer lens to filter the reflection. However, it works well only when the reflection incident angles are close to the Brewster angle [1]. Alternatively, the reflection can also be removed using image processing approaches. In fact, reflection removal is a special topic of the blind image separation (BIS) problem. For the problem of reflection removal, a reflection scene $I_R$ is superimposed onto a background scene $I_B$. The resulting scene $I$ can be described by an additive model as follows:

$$I = I_B + I_R. \qquad (1)$$

Separating $I_B$ and $I_R$ from $I$ is severely ill-posed since we have two unknowns but only one equation. Despite the difficulty of the problem, many solutions were developed in the past 20 years [2-16] and claimed to give good performance. The most typical approaches are based on the independent component analysis [2-4]. These approaches extend the earlier works in blind signal separation to the two-dimensional space. They assume the images are statistically independent and are mixed in different ways such that multiple mixtures of the images can be captured. Such assumption and requirement are difficult to fulfill in general imaging applications. Another type of BIS method is based on the sparse representation of images [5, 6]. They require only a single observation of the scene but assume that the different components in the image have different morphological structures. Such assumption is difficult to achieve for most reflection removal problems since in most cases the morphological structures of the background and reflection images are the same. So rather than just relying on the images' sparse representations, different priors were added to constrain the optimization process. Most priors that the previous methods adopted are gradient based, such as gradient sparsity and gradient independence [7-10]. The former one is a well-known property of natural images; and the latter one is based on the observation that the strong gradients (such as edges) of the background and reflection images are normally non-overlapped. However, the effect of just adding these priors in the optimization process is still limited due to the huge variety of natural images.

Quite recently, the learning-based approaches, such as deep neural networks (DNN), are also applied to the problem of reflection removal [11, 12]. Both the approaches in [11] and [12] claim to achieve reflection removal using only a single observed image. To accomplish this, a number of priors need to be applied to constrain the problem. However, none of these priors is strong enough to clearly distinguish the background and reflection in the image. For instance, [11] assumes the reflection is blurrier than the background and trains two convolutional neural networks (CNNs) to exploit this property. The same assumption is also used in [12] although the high-level features of images are also taken into account when training the CNN. Indeed, the assumption that the reflection is blurry is often invalid, as it is also pointed out in [12]. It introduces the robustness problem to both approaches in general applications.

With only a single image, the reflection removal problem is too unconstrained to solve. Researchers also considered using multiple images. Comparing to single-image based methods, multiple-image based solutions often show better performance. In fact, recent advance in digital imaging technology has allowed multiple images of a scene to be captured conveniently for general users. By taking multiple images of the scene from different viewing angles, we can obtain the depth information of the scene, which is a very useful cue to identify the background and reflection since it is rare to have two uncorrelated images having the same depth. In [8], the different homographies of the background and reflection are adopted for their separation. When the multiple images are aligned to the same view and combined using the background homography, the reflection will be largely misaligned and become blurred. Then a low rank decomposition method is applied to remove the reflection from the combined image. However, the method will work only when the background can be approximated as a plane (such as a painting or a scene far away from the camera). If the background also has a depth range, some parts of it will also be misaligned and removed. Another multiple-image approach using the depth information is described in [9]. In that method, the scale-invariant feature transform (SIFT) flow is used to register the dominant background edges. Since the reflection will fail to register due to its weak intensity, their edges can be separated according to the extent of alignment. However, due to the weak reflection assumption, the method will fail if the reflection is not particularly weak, which is hard to avoid in real scenes. In [10], the difference of the optical flows of the background and reflection images are adopted for their separation. This method however can easily fall into the local minimum because many variables need to be regularized simultaneously. An accurate initial guess is needed to guide the optimization process to the desired solution. Recently, light field cameras are also adopted in the reflection removal problem [13, 14] since image depth can be obtained from light field images effectively. However, the approach in [13] has stringent requirements on the orientation of the camera. Furthermore, both approaches in [13] and [14] assume that the background and reflection have absolutely different depths. Due to the error in depth estimation, these methods often fail to perform when the depths of the background and reflection are close to each other. It will be even worse if some of the components of the background and reflection share the same depth range, which indeed frequently happens in practical situations. For this reason, we recently proposed a reflection removal algorithm [15, 16] based on a new model as shown in Fig. 1. The model assumes a shared depth region of the background and reflection thus allows some of the components in the background and reflection scene to have the same depth values. The new model leads to a more robust algorithm. However, due to the massive optimization processes, the algorithm is rather time consuming which hinders its practical application.

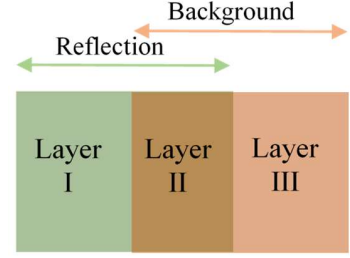

Fig. 1. The model adopted by the proposed algorithm. In this model, Layer I and III comprise the edges of the reflection and background images respectively with distinct depth values. Layer II is shared by the background and reflection images. It contains their edge pixels with similar depth values.

In this paper, a new multiple-image based reflection removal algorithm using deep neural networks is proposed. For the proposed algorithm, multiple images of a scene are captured to evaluate the depth of the image edges using a CNN. Based on the edge independence property, the new algorithm classifies the background and reflection edges based on their depth value. The proposed algorithm also follows the model in Fig. 1 that it allows the background and reflection edges to have a shared depth range (layer II in Fig. 1). The edge pixels in the shared depth region, which are error prone, are not used in the reflection removal process. Rather, they are regenerated based on those in layer I and layer III in Fig. 1 which we have higher confidence about their classes (reflection or background). The original reflection removal problem is thus converted to become an edge regeneration problem. To regenerate the missing edges (in the shared depth region) from the classified edges, we adopt a Wasserstein generative adversarial network (WGAN) [17]. Recently, GANs [18] has drawn much attentions from researchers due to their strong ability in generating new samples following the statistical characteristics of the training dataset. Due to such property, GANs have been successfully applied to various inverse problems, like super-resolution [19], inpainting [20] and denoising [21]. A GAN contains a generator producing new samples and a discriminator jointly trained to evaluate the difference of the generated samples from the real samples. The goal is to train a generator that can synthesize new samples following similar distributions of the ground truths such that they cannot be distinguished by the discriminator. From an analysis of the images with reflection, we notice that there is a significant difference between their statistical distribution and that of normal images. A GAN can learn through the massive training process to generate the required data that follow the distribution of normal images to regenerate the image edges in the shared depth region. However, the training of GAN is a minimax process, which can be unstable and difficult to converge. To conquer this difficulty, we adopt WGAN which applies the Wasserstein distance to its loss function [17]. The training of WGAN is much faster and can converge in a more stable manner than the original GAN. Based on the estimated background edges, we use another auto-encoder to extract the background image from the original one. The flowchart of the entire algorithm is shown in Fig. 2.

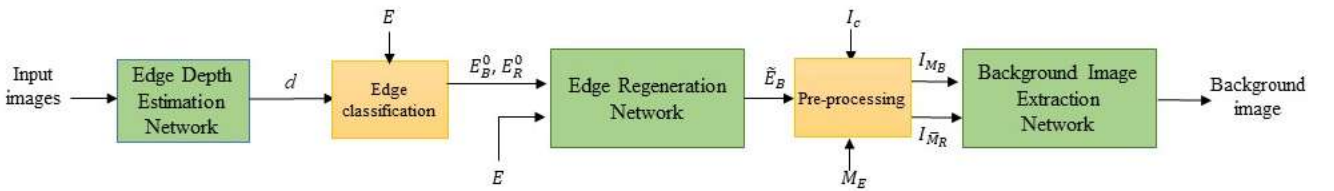

Fig. 2. The flowchart of the entire framework.

To summarize, our main contributions of this paper are as follows:

*1)* We propose a novel deep learning based algorithm to solve the ill-posed reflection removal problem. This approach has no pre-requisites as in the previous approaches on the property of the reflection image (such as blurriness [11, 12] or weak intensity [9]). It also does not have the restrictions as in the previous approaches due to the use of different motion models [8-10] and the need of accurate initialization of the parameters [10].

*2)* For images with reflection, their depth is known to be ambiguous in general. We develop a new CNN approach that directly estimate from the multiple-image input the depth values along the image edges. Due to the edge independence property, we can classify many of the edge pixels as belonging to the background or reflection without ambiguity.

*3)* Since the edges in the shared depth region are error prone, we regenerate these edges by using a WGAN. Thus, we convert the original reflection removal problem to be an edge regeneration problem. We also show that an image with reflection has a significant difference in its statistical distribution from that of normal images. A WGAN can learn through the massive training process to generate the required data that follow the distribution of normal images.

*4)* Instead of using the traditional time-consuming optimization process, we use another auto-encoder to extract the background image from the original one based on the estimated background edges. High level features are adopted in the training of the auto-encoder and the input is also pre-processed to remove the strong edges of reflection. They both contribute to the improved performance of the proposed algorithm.

*5)* Since DNN techniques are adopted in all components of the proposed algorithm, a significant improvement in the computation speed is achieved as compared to the traditional optimization approaches.

The rest of the paper is organized as follows: after the introduction in Section I, we describe the CNN we used to estimate the depth values of the image edges in Section II. In Section III, we present the WGAN which is used to regenerate the missing background edges. In Section IV, we introduce the auto-encoder to extract the background image based on the estimated background edges. In Section V, we show the experimental, comparison and ablation analysis results. Finally, we draw the conclusion in Section VI.

## II. ESTIMATING THE DEPTH OF IMAGE EDGES

Image based depth estimation has been extensively studied for a few decades. The main strategy is to match the feature points in stereo image pair or multiple images taken at slightly different viewpoints [22-24]. However, for images with reflection, the background and reflection overlap each other in each pixel. Since usually they have different depths, image patches that are matched in the background images will become unmatched after the reflection images are superimposed. Fig. 3 shows an example that an image pair (Fig. 3(a)) is superimposed with another image pair (Fig. 3(b)). It can be seen that a pair of matched background patches in Fig. 3(a) will have large difference after the reflection images are superimposed in Fig. 3(b). Depth estimation using the patch matching method on images with reflection obviously will have large error. However,

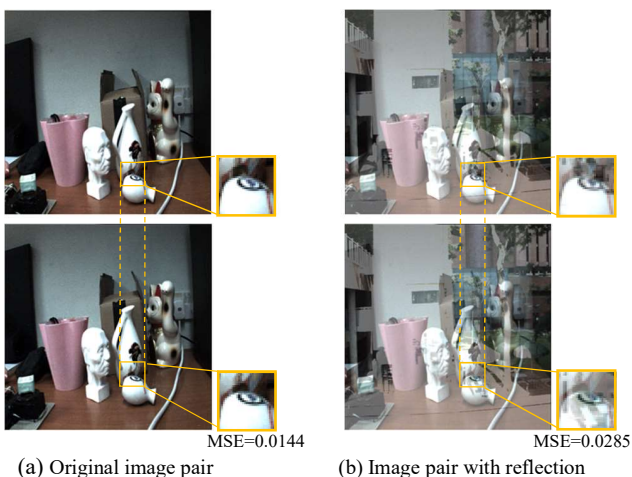

(a) Original image pair      (b) Image pair with reflection

MSE=0.0144        MSE=0.0285

Fig. 3. (a) An image pair of a scene taking at slightly different viewing angles. The boxes show a pair of matched patches in the two images. (b) The image pair in (a) is superimposed by another image pair. The boxes show the same patches of (a) but now have large difference.

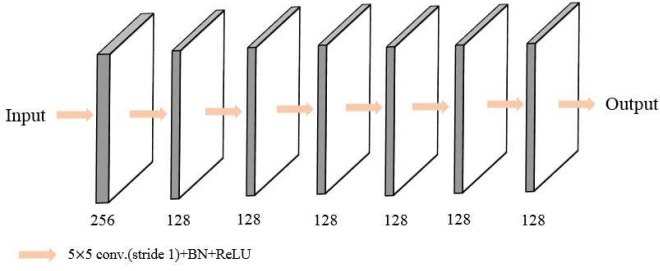

Input ... Output

256  128  128  128  128  128  128

5×5 conv.(stride 1)+BN+ReLU

Fig. 4. The architecture of the Edge Depth Estimation Network.

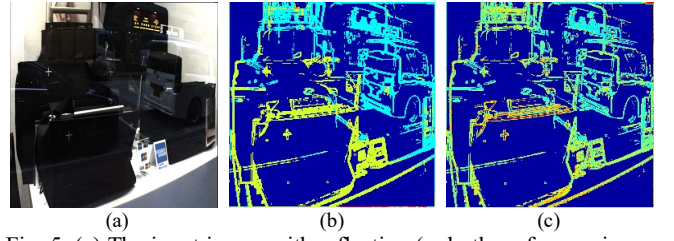

(a)　　　　(b)　　　　(c)

Fig. 5. (a) The input image with reflection (only the reference image is shown); (b) the edge depth estimated using method [27]; (c) edge depth estimated using the proposed network. In (b) and (c), the red and blue colors represent the large and small depth values.

based on the gradient independence property, the strong gradients (such as edges) of the background and reflection seldom overlap each other. Therefore, instead of matching image patches, we estimate the depth by matching only the edges of different images. To achieve this, an Edge Depth Estimation Network is designed using a CNN with architecture as shown in Fig. 4. The network contains eight layers with 256 channels at the beginning, 128 channels in middle six layers and one channel in the last layer for outputting the depth map. The kernel size is 5x5. There are batch normalization layers and ReLU following every convolutional layer except the last one. We assume images from 5 different viewing angles are available. The middle one is selected as the reference image. The subsequent reflection removal procedure will be applied only to the reference image. The other images are for depth estimation. All images are stacked together and fed to the proposed network for estimating the depth along the edges of the reference image. Similar to [25, 26], we train the network by minimizing the loss $\mathcal{L}^D$ between the gradients of the reference image $I_c$ and the other images $I_n$ defined as follows:

$$\mathcal{L}^D = \sum_{n,x} \left\| A_n(x) \cdot I_n(x) - A_n(x) \cdot I_c \left( x + B_{n,c} \cdot d(x) \right) \right\|^2, \quad (2)$$

where $d$ is the depth; $x$ is the pixel coordinate; $n$ is the index of the images; and $B_{n,c}$ is the baseline difference between the reference image and the $n$th image. In (2), $A$ represents the gradient map. Hence, only the image edges are considered in the loss function. Note that in this loss function, we do not need any ground truth depth map. This unsupervised training strategy can avoid the difficulty of collecting the labels of the samples. In the testing phase, the input 5 images are stacked and fed into the network, which directly output the depth map of the edges of the reference image. In Fig. 5, a brief comparison is made between our approach and another CNN based multiple-image depth estimation method [27]. For the ease of visualization, the depth values of only the strong edge points are shown in Fig. 5. Method [27] uses the image patch matching method for depth estimation and we can find many errors in the estimation result. For instance, both the background and reflection are estimated to have the same depths in the top right-hand corner of the image. The error is caused by the aforementioned problem that the patch matching method can have large error when applying to images with reflection. In contrast, the proposed approach which is based on image edge matching gives much higher accuracy as shown in Fig. 5.

The estimated depth values will be used to classify the image edges as belonging to the background or reflection. The background edges are then used to help in extracting the background image. The details are described in the following sections.

## III. EDGES REGENERATION USING WGAN

As mentioned in Section I, background and reflection scenes do not necessarily have different depths. It is frequent that some of their components share the same depth range. Even if they have totally different depth ranges, edges having similar depth values are always difficult to be accurately classified due to the possible error in depth estimation. To solve the problem, we suggest that the edges in the shared depth region, which are error prone, should not be used in the reflection removal process. Rather, we regenerate the background edges in the shared depth region based on those having more distinct depth values. It is achieved by the proposed Edge Regeneration Network as shown in the third functional block in Fig. 2. We first extract the edges in layer I and III of the model in Fig. 1. We apply the $k$-means clustering method as in [16] to obtain two depth thresholds $K_1$ and $K_2$ ($K_2 > K_1$). Details of determining the thresholds can be found in [16]. Without loss of generality, we assume the background has a larger depth range than the reflection. It is just a change of symbols if it is the other way round. Thus, edges with depth values smaller than $K_1$ are classified as the reflection edges (layer I in Fig. 1). Edges with depth values larger than $K_2$ are classified as the background edges (layer III in Fig. 1). Those in between are classified as in the shared depth region (layer II in Fig. 1). The background edges in the shared depth region will be regenerated by using a WGAN. The details will be described in the following subsections.

### A. Wasserstein Generative Adversarial Networks

Compared to CNN which only minimizes the distance in the loss functions, GAN targets to produce samples that are close to the ground truth and cannot be distinguished by a discriminator. The learning process of a GAN can be described by a minimax optimization as follows:

$$\min_G \max_D \mathbb{E}_{x \in \chi}[\log(D(x))] + \mathbb{E}_{z \in Z}\left[\log\left(1 - D(G(z))\right)\right], \quad (3)$$

where $\mathbb{E}$ refers to the expectation operator; $G$ and $D$ represent the generator and discriminator respectively. $G$ is trained to

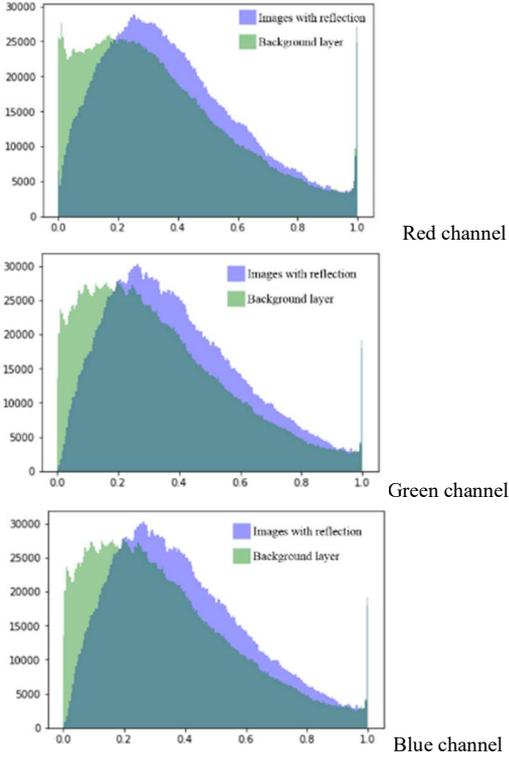

Red channel

Green channel

Blue channel

Fig. 6. Distributions of the edges of normal images (green), and with reflection (blue).

map the input $z$, which follows the distribution $Z$ to the target $x$, which follows another distribution $\chi$. A discriminator $D$ is jointly trained to distinguish the output of $G(z)$ from the real sample $x$ by maximizing a loss function. The goal is to train a generator $G$ which can generate fake data that the discriminator $D$ cannot distinguish. At that time, the output of $G(z)$ should be very close to the target $x$. However, the training of GAN is unstable and difficult to converge. Therefore, we adopt WGAN which inherits the ability of GAN but shows stable and fast convergence. It is achieved by using the Wasserstein distance in its loss function plus a few slight modifications to the training process. The learning process of WGAN can be described as follows:

$$\min_G \max_D \mathbb{E}_{x \in \chi}[D(x)] - \mathbb{E}_{z \in Z}\big[D\big(G(z)\big)\big]. \tag{4}$$

In the training, it also requires to remove the sigmoid activation in the last discriminator layer and clip the weight range of the discriminator to force it to be 1-Lipschitz [17]. With such modifications, we can efficiently train a WGAN to estimate the background edges.

### B. Using WGAN for edge regeneration

We consider WGAN as a suitable tool for this problem because of its strong ability in synthesizing new data following a given statistical distribution. A study was made to understand the difference in distribution between an image with and without reflection. We made use of about 100 real-life images (with reflection) from a dataset [28]. Since the background ground truths are also provided in the dataset, we can compute the histograms of these images with and without reflection. Fig. 6 shows the resulting histograms of only the edge points of these images. It can be seen that there is a significant shift in the bias and skewness for the images with reflection. It is because light is additive. Images with reflection in general have higher intensity than normal images. Such difference in distribution can be fully exploited by WGAN through massive training when generating the edge points of the background from the original image with reflection.

Different from the traditional applications of WGAN (such as image-to-image translation, etc.) that a large degree of freedom is allowed for the image generated, an effective reflection removal algorithm is expected to give an output as close to the background image as possible. The problem is more like an inverse problem than image generation. To enhance the ability of WGAN in generating the missing edges of the background, we suggest a structure similar to the conditional GAN [29]. Rather than just fooling the discriminator, the objective of the generator is revised to also ensuring the output is close to the ground truth background edges $E_B$ in an L2 sense. Hence, when the results of the discriminators are incorporated, it forms a typical regularized minimization process as follows:

$$\tilde{E}_B = \min_{G_B}\|G_B(z) - E_B\|_2^2$$
$$- \lambda_1\left(D_B\big(G_B(z)\big) + D_R\big(E - G_B(z)\big)\right), \tag{5}$$

where $E$ is the image edges; $E_B$ is the ground truth background edges and $\lambda_1$ is the Lagrange multiplier to balance the two terms in (5). Note that the discriminator $D_B$ is trained to give a large value if $G_B(z)$ gives an output close to the ground truth background edges and vice versa, whilst the discriminator $D_R$ is trained to give a large value if $E - G_B(z)$ gives an output close to the ground truth reflection edges and vice versa. So, they combine to form a prior of $G_B(z)$ to regularize the first term in (5), which is a typical approach used in the inverse problem. Note that if $G_B(z)$ gives the background edges, $E - G_B(z)$ will give the reflection edges. Thus, the term $D_R\big(E - G_B(z)\big)$ strengthens the prior with the discriminator $D_R$.

The architectures of the generator and discriminator are shown in Fig. 7. The structure of the generator $G_B$ is similar to a U-net [30] with skip connections, which has been shown to be effective in the inverse problem in [20]. The two discriminators $D_B$ and $D_R$ have the same CNN structures with 6 downsampling layers as shown in Fig. 7. The training of the networks can be performed in an iterative manner as in the original WGAN. For each training image, an input vector $z$ is formed by stacking its edges $E$ , the initially estimated background edges $E_B^0$ and reflection edges $E_R^0$ (that is, the edges in layers III and I in Fig. 1, respectively). They are sent to the proposed WGAN to guide the generator and discriminators to generate the background edges in the shared depth region following the distribution in the background image. Assume that $m$ samples of the background and reflection image ground truths and $m$ samples of the training images are obtained. We can train the generator $G_B$ by updating its parameters using the gradient descent method to minimize the following function:

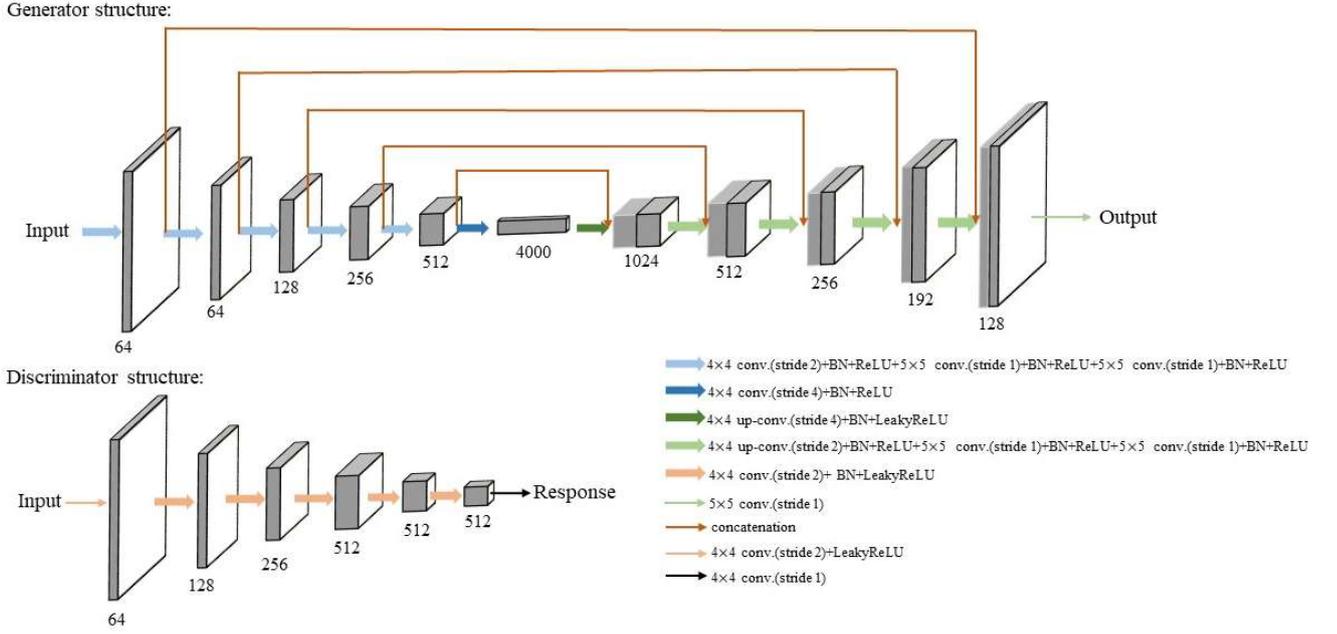

Fig. 7. The network structures of the generator and discriminators.

$$\sum_{i=1}^{m} \|G_B(z(i)) - E_B(i)\|_2^2$$
$$- \lambda_1 \Big( D_B \big( G_B(z(i)) \big) \qquad (6)$$
$$+ D_R \big( E - G_B(z(i)) \big) \Big).$$

The trained generator $G_B$ is then used to train the discriminators $D_B$ and $D_R$. Assume another $m$ samples of the background and reflection image ground truths and $m$ samples of the training images are obtained, the parameters of $D_B$ and $D_R$ are updated using the gradient ascent method (plus weight clipping) to maximize the following cost functions:

$$\sum_{i=1}^{m} D_B(E_B(i)) - D_B\big(G_B(z(i))\big); \qquad (7)$$

$$\sum_{i=1}^{m} D_R(E_R(i)) - D_R\big(E - G_B(z(i))\big). \qquad (8)$$

In (8), $E_R$ is the ground truth reflection edges, which can be derived from the ground truth reflection image. The trained $D_B$ and $D_R$ are then used to train $G_B$ again to obtain a better generator. The process repeats until converged. In the testing phase, the input $z$, which comprises the edges $E$ of the input image, the initial background edges $E_B^0$ and reflection edges $E_R^0$, is fed to the trained generator $G_B$ to obtain an estimate of the background edges $\tilde{E}_B$. The reflection edges $\tilde{E}_R$ can also be obtained by masking out $\tilde{E}_B$ from $E$. A background binary edge map $\tilde{M}_B$ is also obtained by selecting edges in $\tilde{E}_B$ whose gradient magnitudes are above a threshold $\sigma$. We empirically set $\sigma = 0.05$ in our experiment. We then feed $\tilde{E}_B$ and $\tilde{M}_B$ to the Background Image Extraction Network as shown in Fig. 2 to obtain the reflection-free background image. Fig. 8 shows an example of $\tilde{M}_B$ obtained from the proposed Edge Regenerating Network. We denote the binary edge masks for $E$ and $E_B^0$ as $M_E$ and $M_B^0$ respectively. We can see that $M_E$ contains both the background and reflection edges, while the initial background edge map $M_B^0$ only contains a portion of the background edges. As shown in Fig. 8(d), the proposed Edge Regeneration Network successfully regenerates a large amount of missing background edges while ignoring most of the reflection edges.

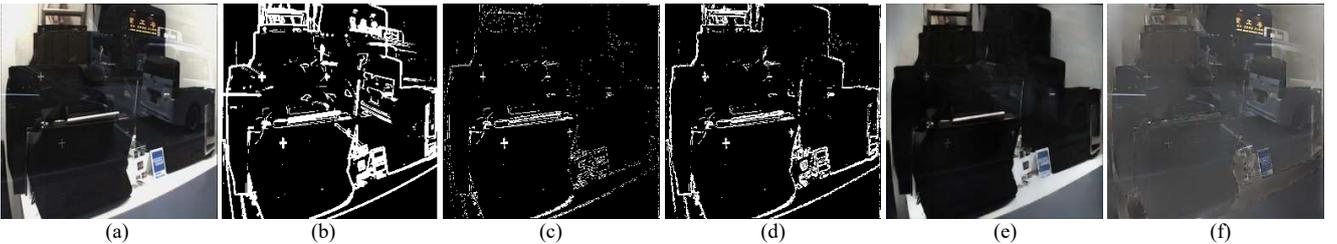

Fig. 8. (a) The input image with reflection. (b) The binary edge map $M_E$. (c) The initial background edge map $M_B^0$. (d) The estimated background edge map using the proposed Edge Regeneration Network. (e) The extracted background image based on the edge map in (d). (f) The reflection obtained by deducting the background image from the input image. The mean of (f) is adjusted to the input image for the ease of visualization.

## IV. Background Image Extraction Based On Edges

In [15-16], we have demonstrated that, by using the traditional optimization method, we can extract the background image from the original one (with reflection) if the edges of the background are given. However, the required iterative process is very time-consuming. To be compatible with the other components of the algorithm, we consider also using the highly efficient DNN for this problem.

To the best of our knowledge, there are very few DNN approaches for extracting the background images based on their edges. The only one we are aware of is the I-CNN in the method CEILNet [11]. However, the performance of I-CNN is rather unstable that the resulting image can lose many background details while keeping the reflection residual. It is because I-CNN works based on the assumption that the reflection is blurry. When the image contains reflection with strong edges, it is difficult for I-CNN to totally remove them. To solve the problem, we develop a new Background Image Extraction Network, which has an auto-encoder structure the same as that in Fig. 7 (upper). To remove the strong edges of the reflection remained in the resulting image, we pre-process the input image by removing the reflection edges. To do so, we first compute from the estimated background edge map $\widetilde{M}_B$ a residual map $\widetilde{M}_{\bar{B}} = M_E - \widetilde{M}_B$, which mainly indicates the positions of the reflection edges. Then, we obtain an image $I_{\bar{M}_R} = (I_c - I_c \cdot \widetilde{M}_{\bar{B}})$, which is the original reference image without the reflection edges. Both $I_{\bar{M}_R}$ and the background edges $I_{M_B} = (I_c \cdot \widetilde{M}_B)$ are stacked as the input signal $z$ and fed to the proposed Background Image Extraction Network. For training the network, we first use an L2 norm loss in (9) to confine the resulting image to follow the ground truth background at pixel level,

$$\mathcal{L}_{rec}^I = \|G^I(z) - I_B\|_2^2, \tag{9}$$

where $I_B$ is the ground truth background image and $G^I(z)$ is the network output given the input $z$. In addition, we add a perceptual loss in (10) to ensure the resulting image to follow the human perception,

$$\mathcal{L}_p^I = \|V(G^I(z)) - V(I_B)\|_2^2, \tag{10}$$

where $V$ represents the output of the 14th layer of the pre-trained VGG-16 network. Using the intermediate responses of high-level network features is an effective way to measure the perceptual similarity [31]. Thus, the following overall loss function is used to train the proposed Background Image Extraction Network:

$$\mathcal{L}^I = \mathcal{L}_{rec}^I + \lambda_2 \mathcal{L}_p^I, \tag{11}$$

where $\lambda_2$ is used to balance the two loss functions. At the testing phase, the estimated background edge map $\widetilde{M}_B$ obtained from the Edge Regeneration Network and the reference image $I_c$ are used to obtain $I_{\bar{M}_R}$ and $I_{M_B}$. They are then stacked together and fed to the Background Image Extraction Network. Fig. 8(e) and (f) show an example of extraction result and its residual (reflection layer) using the estimated background edge map in Fig. 8(d). We can see that the proposed network successfully extracts the background image from the original one based on the background edge map.

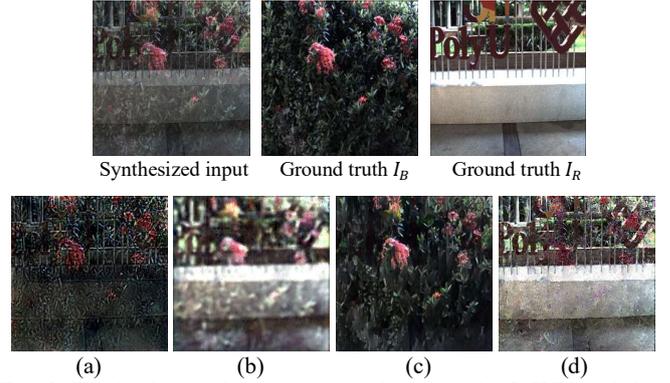

Fig. 9. The background extraction results using the I-CNN and the proposed Background Image Extraction Network. (a) and (b) are the extracted background and its residual, respectively, using *the Edge Depth Estimation Network + Edge Regeneration Network + I-CNN*. (c) and (d) are the extracted background and its residual, respectively, using *the Edge Depth Estimation Network + Edge Regeneration Network + the proposed Background Image Extraction Network*.

A brief comparison with I-CNN is shown in Fig. 9. To isolate the performance in background extraction, both the proposed Background Image Extraction Network and I-CNN use the estimated background edge map generated by the Edge Regeneration Network. Since the reflection is not particularly blurred in the synthesized image, we can see that the strong edges of the reflection remain in the result of I-CNN. We also notice that many background details are missing. The proposed Background Image Extraction Network can well recover the background components while removing the reflection, since there is no assumption about the blurriness of the reflection and we also incorporate the human perception in the training process. More detailed comparisons can be found in Section V.

## V. Experiments and Evaluation

For evaluating the performance of the proposed approach, we compare it with the state-of-the-art methods both quantitatively and qualitatively.

### A. Training details

We assume that five images of slightly different viewing angles are available as the input of the proposed Edge Depth Estimation Network. For convenience, we obtain the required images for the training of the network by using a light field (LF) camera, which can directly capture array images of the target scene in a single shot. We extract five of the captured images and input them to the network. For quantitative evaluation, we synthesize the required training images with reflection by randomly adding two sets of LF images together with different weights. More specifically, we capture 318 sets of LF images and resize them to $256 \times 256$ pixels. They are randomly added together and finally 112,225 images with reflection are synthesized as the training samples. To further increase the training samples, we augment the data by cropping the images into many $128 \times 128$ patches at every interval of 16 pixels, then randomly flipping and rotating them at every 90 degrees. The Edge Depth Estimation Network is trained using the ADAM solver [32] with learning rate $2 \times 10^{-5}$, $\beta_1 = 0.9$ and $\beta_2 = 0.999$. For both the Edge Regeneration Network and

| Method | PSNR of the recovered background |
|---|---|
| Synthetic input | 13.094 |
| LS-SIFTF [9] | 18.912 |
| SID [8] | 15.488 |
| LS-DS [13] | 18.855 |
| CEILNet [11] | 17.714 |
| PLNet [12] | 19.092 |
| Proposed w/o Edge Regeneration | 22.774 |
| Proposed w/o discriminators | 23.224 |
| Proposed w/o $I_{\bar{M}_R}$ | 23.340 |
| Proposed w/o $I_{M_B}$ | 23.220 |
| Proposed | **24.031** |

Background Image Extraction Network, we only use the flipped and rotated images to augment the dataset. It is because a cropped patch may not have sufficient amount of edges for training, as edges are sparse in nature. Similar to [17], we use the RMSprop solver [33] to train the generator and the discriminators of the Edge Regeneration Network with learning rates $2 \times 10^{-4}$ and $2 \times 10^{-5}$ respectively. For the Background Image Extraction Network, we also use the RMSprop solver [33] with learning rate $2 \times 10^{-4}$ for its training. The parameters $\lambda_1$ and $\lambda_2$ are set as $2.5 \times 10^{-3}$ and 1.25 respectively. The training and testing are both performed on a computer with Core i7 7820X CPU using a GTX 1080 Ti.

### B. Quantitative evaluation

A quantitative comparison is made between the proposed algorithm and a few recent methods, including the traditional optimization based approaches such as SID [8], LS-SIFTF [9], and LS-DS [13]; as well as two CNN-based methods CEILNet [11] and PLNet [12]. Except LS-DS which is implemented by us according to their paper, other methods are implemented by the source codes published in their websites. Because LS-SIFTF and SID require relatively large disparities between images, the images we captured using the LF camera with small baselines cannot be directly used to test these two approaches. To solve the problem, we put the LF camera on a tripod and

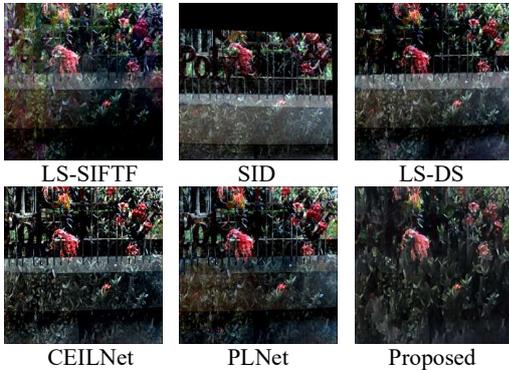

LS-SIFTF    SID    LS-DS

CEILNet    PLNet    Proposed

Fig. 10. The background extraction results using different approaches. The original images and the synthesized image are shown in Fig. 9 (top row).

shift the camera up and down to five preset heights. For each height, we capture one set of LF image for each scene. Using only the central view of each LF image, we can obtain, for each scene, five images of relatively large disparities. We capture 20 groups of such images of different scenes and create ten groups of images with reflection by adding ten of them to the other ten with the weights 0.6 and 0.4. These images are used to test the LS-SIFTF and SID methods. On the other hand, the method LS-DS requires LF images as input. For each group of LF image captured, this time we just use one of them for each scene. We extract the central $5 \times 5$ images of each LF image so that we have twenty $5 \times 5$ images. They are mixed with a similar method as mentioned above to form ten testing images (with reflection) for LS-DS. CEILNet and PLNet are single-image based reflection removal methods, thus we directly input the central view of each LF image to test these networks. Because LS-SIFTF, SID, LS-DS can only perform well with relatively higher resolution images, we feed images with resolution $625 \times 434$ to those methods and resize their results to $256 \times 256$ pixels for comparison. CEILNet and PLNet are directly fed with images with size $256 \times 256$ pixels. Fig. 10 shows one of the comparison results based on the testing images mentioned above. It can be seen that the proposed algorithm gives the best result compared to other methods. The average PSNRs of all the testing algorithms are shown in Table I. Because the results of LS-SIFTF and SID can have large bias in the mean value which can give very low PSNRs, we normalize the mean values of all the results to be the same as the ground truths. As shown in Table I, the proposed method significantly outperforms the other competing methods. It is because all other methods have different assumptions about the input images. For instance, LS-SIFTF requires the gradients of the background to be much larger than the reflection; SID requires the background to be planar; LS-DS requires the background and reflection to be at different sides of the focal plane and the normal line of the camera must be perpendicular to the scene; CEILNet and PLNet have a stringent assumption that reflection must be blurry. They all introduce the errors to the reflection removal process in case the input images do not follow exactly the respective assumptions.

### C. Ablation analysis

To understand the effectiveness of different components of the proposed networks, an ablation analysis is carried out based on the quantitative evaluation above. First, we investigate the importance of the proposed Edge Regeneration Network. As mentioned above, the objective of the Edge Regeneration Network is to regenerate the background edges of which the depth values are similar to those of the reflection. Without the network, we can only classify them to be either belonging to the background or reflection. To do so, we use the same $k$-means clustering method as in [16] but only two clusters are generated for the classification of every edge pixel based on its depth. The background edges are then fed to the Background Image Extraction Network. As shown in Table I, the average PSNR of the recovered background images is reduced by about 1.3 dB without the Edge Regeneration Network. It shows the importance of the network to the proposed reflection removal algorithm.

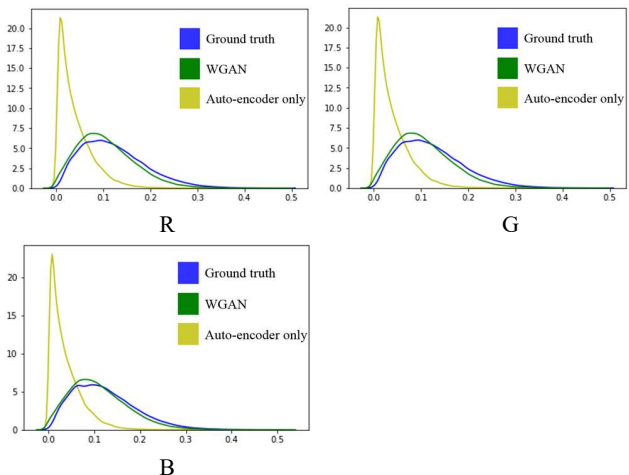

Fig. 11. The fitted distributions of the estimated background edges $\tilde{E}_B$ given by the proposed WGAN and only the generator (trained without the discriminators). R, G, and B refer to the red, green and blue channels of the image, respectively.

Second, we show how the discriminators in the Edge Regeneration Network contribute to the estimation of the background edges $\tilde{E}_B$. We compare using the proposed WGAN (consists of the generator jointly trained with the discriminators using the loss functions (6) to (8)) and only the generator (trained without the discriminators) for estimating $\tilde{E}_B$ of the above images used in the quantitative evaluation. Fig. 11 shows the fitted distributions of the estimated background edges using the proposed WGAN and only the generator. We can see that the background edges given by WGAN has the distribution very close to that of the ground truth. It is because WGAN tends to constrain the generated samples following the distributions of the ground truths, such that the discriminators cannot distinguish them from the real ones. Without the discriminators, the generator can only give the estimation in the mean square sense, which is known to have problem in dealing with the transients in the data (such as image edges). Table I also shows that without the discriminators, the average PSNR of the recovered background images decreases by about 0.8 dB. It shows the contribution of the discriminators to the proposed algorithm.

Finally, we also evaluate the importance of the input terms $I_{M_B}$, $I_{\bar{M}_R}$ for the Background Image Extraction Network. As mentioned in Section IV, $I_{M_B}$ is the estimated background edges while $I_{\bar{M}_R}$ is the original image with the estimated reflection edges removed. Although they provide related information to the Background Image Extraction Network to recover the background image, they have to be input together to the network to achieve the best performance. Without any one of them, the overall PSNR will decrease by about 0.7 dB.

### D. Qualitative evaluation

For qualitative evaluation, we compare the visual quality of the extracted background and reflection images using different methods. In this evaluation, the testing images are directly captured in real-life environment, such as in front of a glass, etc., so that reflection of unwanted scene is added to the image. Since we do not have the ground truth background of these images, we can only evaluate the performance by visual inspection. In Fig. 12, we show the results using different methods. As shown in the figure, LS-SIFTF [9] cannot correctly separate the reflections from the backgrounds when both of them have strong gradients. The performance of it is only relatively reasonable for the fourth scene whose reflection is relatively weak. However, there are still many residuals remained in the regions with strong reflections. For SID [8], it assumes the background layer is planar and uses the homography to register the background while blurring the reflection. Thus, it can only deal with planar background scenes. In fact, even the background is planar, the features of the reflection can affect the homography estimation. Therefore, we can see that the resulting images are blurry due to inaccurate registration. For method [13], it requires the background and reflection to have an absolutely different depth ranges and it also requires the camera to be perpendicular to the target scene. Such stringent requirements to the pose and photography environment introduce much difficulty to remove the reflection in practice. For CEILNET [11] and PLNet [12], they assume the reflection is much smoother than the background. They fail to remove the strong and sharp reflection components in the images. Without the abovementioned limitations, the proposed method successfully extracts the backgrounds and separates the reflections for all scenes as shown in Fig. 12.

### E. Running time

We also compare the running time of different testing methods by taking the average processing times of these methods on five real-life images. This time we only evaluate the computational cost regardless of performance. Therefore, we feed images with size $256 \times 256$ to all methods. The results are shown in Table II. It can be seen that traditional optimization-based methods such as [8], [9], [13] and [16] use much longer times compared to the DNN based methods, such as CEILNET, PLNet and the proposed one. It is because those optimization-based methods require iterative operations on huge matrices, which can take very long time. In contrast, the feed-forward architectures of different DNN approaches efficiently utilizes the massive parallel structure of GPU. They can complete the whole process within only one second.

TABLE II.
THE AVERAGE EXECUTION TIMES OF DIFFERENT METHODS

| Method | Average Time |
|---|---|
| LS-SIFTF [9] | 130.59 s |
| SID [8] | 58.95 s |
| LS-DS [13] | 17.01 s |
| LS-LFGS [16] | 69.51s |
| CEILNet [11] | 0.82 s |
| PLNet [12] | 1.15 s |
| Proposed | 0.88 s |

## VI. CONCLUSION

In this paper, we propose a novel approach to solve the reflection removal problem. The proposed approach fully utilizes the deep neural network techniques in estimating the edge depths, regenerating the edges with potential classification errors, as well as extracting the background image based on its edges. The proposed method does not have any pre-requisite requirements on the intensity or smoothness of the reflection image. It also avoids the limitations due to the motion models used in the traditional multiple-image based reflection removal methods. The proposed approach shows superior performance over the state-of-the-art methods. It also shows significant improvement on the computation speed compared to the existing approaches using traditional optimization methods. The current networks are only trained with synthetic images. We would expect the performance can be further improved if the networks can be trained with real-life images with reflection. However, extra measures would be required to label these images. It will be one of our future works.

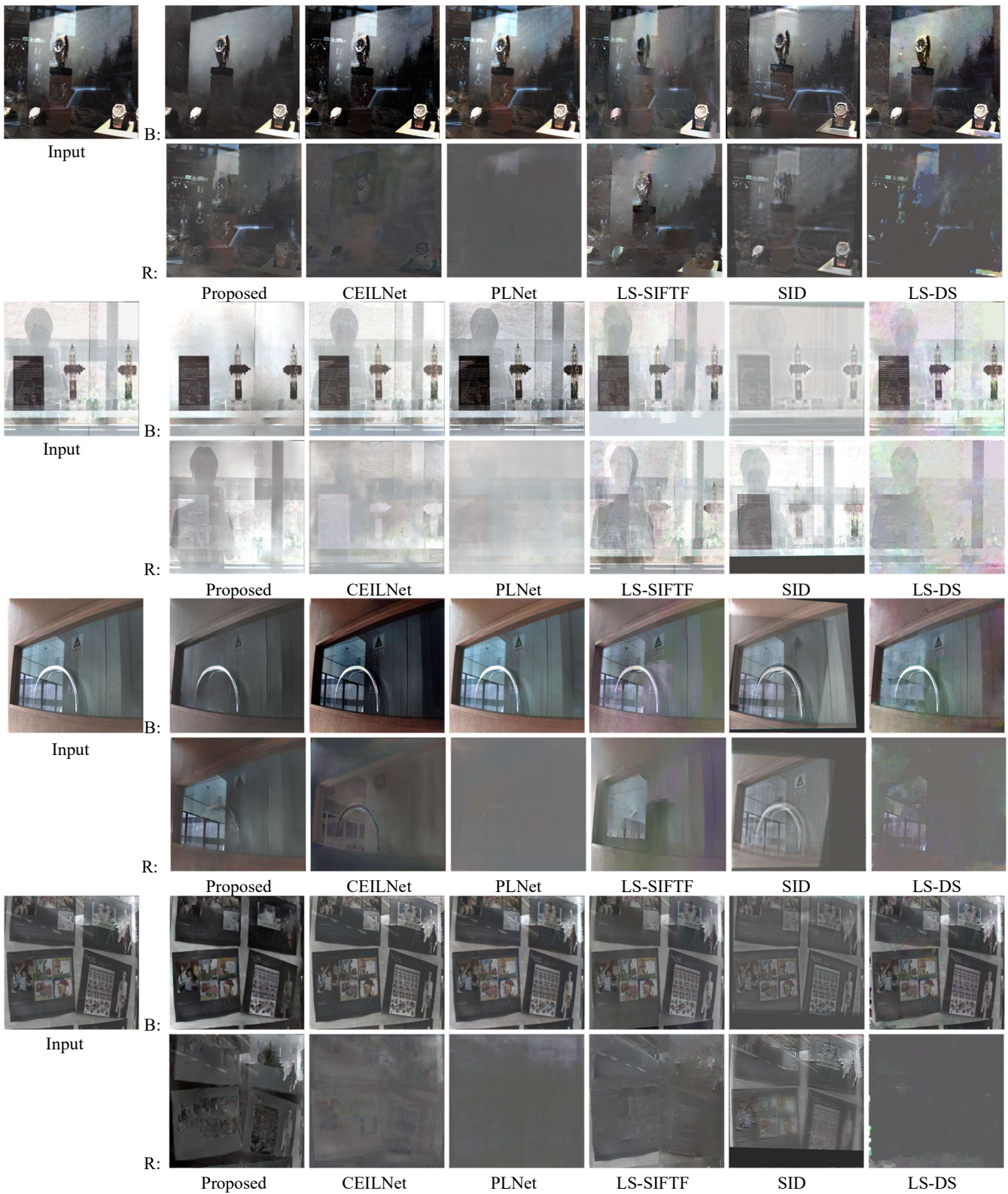

Fig. 12. The comparison of different methods on real-life images with reflection. We adjust the mean values of the results to the input image for the ease of visualization. 'B' and 'R' represent the background and reflection respectively.

### ACKNOWLEDGEMENT

This work is supported by the Hong Kong Polytechnic University under research grants RU9P and 4-ZZHM. Besides, we would like to thank the ROSE Lab at the Nanyang Technological University, Singapore, for allowing us to use their 'SIR2' dataset.